# Unveiling the origin of diffusion suppression of hydrogen isotopes at the α-Al$_2$O$_3$(0001)/α-Cr$_2$O$_3$(0001) interfaces

Yuji Kunisada,* Ryotaro Sano, Norihito Sakaguchi

Center for Advanced Research of Energy and Materials, Faculty of Engineering, Hokkaido University, Kita 13 Nishi 8 Kita-ku, Sapporo, Hokkaido 060-8628, Japan.

*Email: kunisada@eng.hokudai.ac.jp

**Abstracts**

It has been reported that the α-Al$_2$O$_3$, a promising tritium permeation barrier material for a fusion reactor, can be grown at low temperatures on the α-Cr$_2$O$_3$ template, and that α-Al$_2$O$_3$/α-Cr$_2$O$_3$ composite films have more efficiently suppress the hydrogen isotope permeation than the single α-Al$_2$O$_3$ film. In this study, we investigated the diffusion properties of hydrogen isotopes at the α-Al$_2$O$_3$(0001)/α-Cr$_2$O$_3$(0001) interfaces using first-principles calculations based on density functional theory. In the α-Al$_2$O$_3$ region near the interfaces, O-H covalent bonds, which are not observed in the bulk α-Al$_2$O$_3$, are formed, and hydrogen isotopes become stable. Such chemical bonds induced by the interfaces are the origin of hydrogen isotope trapping and result in a larger diffusion barrier than in the α-Al$_2$O$_3$ and the α-Cr$_2$O$_3$. It was also found that the suppression of hydrogen isotope diffusion does not occur at the interface site but at sites adjacent to the interfaces. In addition, the interface enhances the oxygen vacancies, which may also suppress hydrogen isotope permeation.

**Highlights**

- In the α-Al$_2$O$_3$ region near the interfaces, O-H covalent bonds, which are not observed in the bulk α-Al$_2$O$_3$, are formed.
- In the cases of tritium, the diffusion barriers are improved by 0.23 eV due to the effect of the interfaces.
- The suppression of hydrogen isotope diffusion does not occur at the interface site but at sites adjacent to the interfaces.

# 1. Introduction

A deuterium-tritium (D-T) nuclear fusion reactor is one of the most promising fundamental solutions to energy and environmental problems because deuterium atoms, the fuel, are abundant in seawater and do not emit carbon dioxide [1,2]. In a fusion reactor blanket, tritium atoms can also be produced by the nuclear reaction of Li atoms with high-energy neutrons from the D-T fusion reaction. However, hydrogen isotopes readily permeate metallic materials, resulting in tritium leakage, i.e., fuel loss and radiation damage to living organisms [3]. In addition, hydrogen isotopes cause a ductile-to-brittle transition, i.e., hydrogen embrittlement, in metallic materials. Hydrogen isotope permeation barriers, such as coatings or liner materials, are widely employed to suppress hydrogen isotope penetration into metallic materials across various hydrogen-related applications.[4,5] In fuel cell vehicles, the recent high-pressure hydrogen storage tanks, referred to as Type IV tanks, operate at hydrogen pressures of 70 MPa or higher and utilize high-density polymers as hydrogen barrier liner materials [6]. When applied in nuclear fusion reactors, these barrier materials must withstand extreme operating conditions, including high temperatures, mechanical stress, highly corrosive environments, and neutron irradiation. Consequently, tritium permeation barriers (TPB) for blanket structural materials have to not only effectively suppress hydrogen permeability but also exhibit excellent chemical and mechanical stability. Key requirements include strong adhesion to structural materials at high temperatures, low thermal expansion mismatch, high corrosion resistance against lithium alloys, and electrical insulators to prevent magnetohydrodynamic (MHD) pressure losses [7]. Therefore, thin ceramic coatings with $Al_2O_3$ [8,9], $Er_2O_3$ [10,11], $TiO_2$ [12], and $Cr_2O_3$ [13-15] have been studied as TPB for decade. $Al_2O_3$ has received considerable attention as a surface coating material due to its high melting point and superior tritium permeation reduction compared to other ceramic materials. The theoretical permeation reduction factor (PRF) of α-$Al_2O_3$ is very high, $10^6$, but stability and PRF decrease as crystallinity decreases.[16] Therefore, it is necessary to create dense crystalline α-$Al_2O_3$ film rather than amorphous $Al_2O_3$ film. However, to obtain the α-$Al_2O_3$ film, high temperatures for deposition and annealing, such as above 1173 K, are required, and cracks in the films with high operating temperatures may occur due to the large difference of thermal expansion coefficients between α-$Al_2O_3$ and steel [17,18]. Many efforts have been made to fabricate α-$Al_2O_3$ at low temperatures. Andersson and co-workers reported that α-$Al_2O_3$ films can be epitaxially grown at 553-833 K on α-$Cr_2O_3$ substrates by reactive radio frequency (RF) magnetron sputtering [19]. The PRF of $Al_2O_3$/$Cr_2O_3$ composite films increased from 230 to 544 with an increase in temperature from 823 to 973 K. The PRF values of $Al_2O_3$ and $Cr_2O_3$ films at the same temperature regions were 95-247 and 24-117, respectively [20]. Therefore, the $Al_2O_3$/$Cr_2O_3$ composite membrane can significantly improve the deuterium permeation suppression properties.

The interfaces in materials typically act as hydrogen-trapping sites, as reported in metal/metal[21,22], ceramics/metal[23,24], and ceramics/ceramics [25,26] interfaces. However, the effects of interfaces are not simply comprehended. In the Fe grain boundaries, it has been reported that the stability of hydrogen atoms does not depend simply on the size of the interstitial sites, and hydrogen atoms rather become unstable at Σ3 grain boundaries of fcc Fe than bulk octahedral sites [21]. These results indicate that it is necessary to accurately evaluate the bonding nature to understand the behavior of hydrogen near the interface. Zhang and co-workers have reported the theoretical study of hydrogen behavior at the α-$Al_2O_3$/α-$Cr_2O_3$ interfaces.[27] However, possibly due to the thin slab model, hydrogen atoms are considerably stabilized across the interface than in the bulk α-$Al_2O_3$, and a theoretical study that reduces the surface effects and focuses on the interface effects is needed to discuss thicker TPB films.

In this paper, we investigated hydrogen isotope permeation behaviors at the α-$Al_2O_3$(0001)/α-$Cr_2O_3$(0001) interface using first-principles calculations based on density functional theory (DFT). Firstly, the formation energy and diffusion barrier of a hydrogen atom in α-$Al_2O_3$ and α-$Cr_2O_3$ were investigated. We also constructed the α-$Al_2O_3$(0001)/α-$Cr_2O_3$(0001) interface model without a vacuum layer and investigated the diffusion barriers of hydrogen atoms at the α-$Al_2O_3$(0001)/α-$Cr_2O_3$(0001) interface. We revealed the isotope effects on hydrogen diffusion barriers by considering zero-point vibrational energy. Finally, we discussed the hydrogen-trapping effects of oxygen vacancies.

## 2. Calculation methods

In this study, we investigate the formation energy and diffusion barrier of a single hydrogen atom in the α-$Al_2O_3$, α-$Cr_2O_3$, and at the α-$Al_2O_3$(0001)/α-$Cr_2O_3$(0001) interfaces by using *ab initio* electronic structure calculations based on density functional theory. We used the Vienna Ab Initio Simulation Package (VASP) [28-31], which employs plane waves as basis functions. We adopted the generalized gradient approximation developed by Perdew, Burke, and Ernzerhof (GGA-PBE)[32] as the exchange-correlation functional. We used the Monkhorst-Pack method [33] for k-point sampling for the first Brillouin zone, with a Gaussian smearing $\sigma$ of 0.05 eV. We set the cutoff energy of the plane-wave basis and the number of k-point sampling to converge the energy of the supercells below 0.01 eV/atom, specifically, 500 eV and 2×2×2. In the case of α-$Cr_2O_3$, the conventional GGA-PBE functional was supplemented with the effective on-site Coulomb interaction $U_{eff}$ of 3.5 eV proposed by Dudarev and co-workers.[34] All the atoms were fully relaxed until the force on each atom was less than 0.02 eV/Å. We calculated the diffusion barriers using the climbing image nudged elastic band (CI-NEB) method [35,36] with five intermediate images. We calculated the electron transfer using Bader charge analysis [37-39]. We also calculated the crystal orbital Hamilton population (COHP) analysis using the LOBSTER package [40,41] to evaluate the covalency of electron orbital hybridization. We used Visualization for Electronic and Structure Analysis (VESTA) [42] to visualize the atomic structure and electron density distribution.

The hydrogen formation energy is defined by the following equation:

$$E_\text{f} = E_\text{bulk}^\text{H} - \left(E_\text{bulk} + \frac{1}{2}\mu_{H_2}\right), \tag{1}$$

where $E_\text{bulk}^\text{H}$ and $E_\text{bulk}$ are the total energy of the systems with and without a hydrogen atom. $\mu_{H_2}$ is the total energy of molecular hydrogen in a cube of 20 Å on a side.

Figure 1(a) shows the 1×1×1 supercell with corundum-type atomic structures containing 30 atoms for α-Al$_2$O$_3$. We note that the supercell for α-Cr$_2$O$_3$ is the same as α-Al$_2$O$_3$ except for lattice constants. The calculated equilibrium lattice constants for α-Al$_2$O$_3$ (α-Cr$_2$O$_3$) are $a$=4.83 Å (4.94 Å) and $c$=13.13 Å (13.83 Å), respectively. These values are in good agreement with experiment values in 2.2%.[43,44] We constructed the α-Al$_2$O$_3$/α-Cr$_2$O$_3$ interface model with 60 atoms connecting the most stable cation-terminated (0001) planes for α-Al$_2$O$_3$, as shown in Fig. 1(b). Since there is a lattice mismatch of 2.5% between the α-Al$_2$O$_3$(0001) and α-Cr$_2$O$_3$(0001), the supercell lengths were relaxed after constructing the interface model. The obtained supercell lengths are $a$=4.86 Å and $c$=27.16 Å. The interatomic distance distributions of Al-O (Cr-O) in the interface model are 1.88-2.01 Å (1.93-2.04 Å), and a reduction of the interatomic distance observed in the surface model did not occur.[45] Therefore, no significant atomic distortion in the interface model is observed even after structural relaxation.

## 3. Results and discussion
### 3.1 Diffusion of H atoms in α-Al$_2$O$_3$ and α-Cr$_2$O$_3$

The most stable sites of H atoms in the α-Al$_2$O$_3$ are the center of the largest interstitial sites. The corresponding hydrogen formation energy is 3.51 eV. The difference in the hydrogen formation energy with the results of 2×2×1 supercell is less than 0.04 eV, which indicates that the interaction between H atoms in repeated supercells is negligibly small. Figure 2(a) shows the differential electron density distribution in the vicinity of H atoms in the α-Al$_2$O$_3$. While hydrogen atoms have a slight negative charge of −0.15 $e$ and asymmetric electron distribution toward Al$^{3+}$, there are no regions where electron density significantly increases in the interatomic region near H atoms. Therefore, H atoms in the α-Al$_2$O$_3$ are almost neutral and have no covalent bonds, which results in the high hydrogen formation energy. On the other hand, H atoms in the α-Cr$_2$O$_3$ have covalent bonds with O atoms and show relatively lower hydrogen formation energy of 1.77 eV than that in the α-Al$_2$O$_3$. Figure 2(b) shows the differential electron density distribution in the vicinity of H atoms in the α-Cr$_2$O$_3$. We can see the electron density increase between H and O atoms. O-H bond energy evaluated from COHP analysis is −8.07 eV. These results indicate the formation of O-H covalent bonding. Such a difference in bonding natures in the α-Al$_2$O$_3$ and α-Cr$_2$O$_3$ originates from cation elements.[46] It has been known that a H atom is stable as H$^+$ in insulators and semiconductors when the conduction band minimum of bulk materials is lower than −4.5 eV vs. vacuum level.[47] Conversely, a H atom is stable as H$^-$ when the valence band minimum is higher than −4.5 eV vs. vacuum level. In the present study, the criterion energy level of −4.5 eV vs. vacuum level is located in the band gap of the α-Al$_2$O$_3$ and α-Cr$_2$O$_3$. [48,49] Therefore, H atoms in the α-Al$_2$O$_3$ and α-Cr$_2$O$_3$ have a neutral charge. In the α-Cr$_2$O$_3$, the reduction of Cr$^{3+}$ to Cr$^{2+}$ can lead to the formation of O-H covalent bonds by accepting an excess electron. On the other hand, the instability of Al$^{2+}$ inhibits the formation of O-H covalent bonds in the α-Al$_2$O$_3$.

Figure 3 shows energy profiles through H atom diffusion between the most stable sites in the α-Al$_2$O$_3$ and α-Cr$_2$O$_3$. The diffusion barriers of H atoms in the α-Al$_2$O$_3$ and α-Cr$_2$O$_3$ are 1.14 and 0.92 eV. As shown in Fig. 4(a), at the saddle point in the α-Al$_2$O$_3$, H atoms are located in a small interstitial site with large strain energy and do not form covalent bonds with the other atoms. At the saddle point in the α-Cr$_2$O$_3$, the H atoms are located adjacent to Cr atoms in a small interstitial site, and there is a region of locally increased electron density between H and Cr atoms. The COHP analysis also indicates the formation of the Cr-H covalent bonds with a bond energy of −2.60 eV. These results show that the H atoms can form covalent bonds with the other atoms even at the saddle point. This is because Cr atoms, transition metals, can form covalent bonds with H atoms through localized *d* electrons even after oxidization, which lowers H atom diffusion barriers in the α-Cr$_2$O$_3$.

### 3.2 Diffusion of H atoms at the α-Al$_2$O$_3$(0001)/α-Cr$_2$O$_3$(0001) interfaces

We first investigated the hydrogen formation energies at various interstitial sites near the interfaces to investigate the diffusion properties of H atoms at the α-Al$_2$O$_3$(0001)/α-Cr$_2$O$_3$(0001) interfaces. As shown in Fig. 5, we treated interstitial sites in the vicinity of five O layers, three Al atoms, and three Cr atoms as initial sites for H atoms. The upper and lower interstitial sites of atoms coordinating with hydrogen are named by adding U and L, respectively. Since H atoms are strictly confined in small cages at Al1$^U$, Al2$^L$, Al3$^L$, Cr1$^U$, Cr2$^U$, and Cr3$^L$ sites, we ignored these sites. The initial positions of H atoms are 1.0 Å above and below the O layers, 1.7 Å above and below the Al atoms, and 1.5 Å above and below the Cr atoms. Figure 6 shows the hydrogen formation energy at the α-Al$_2$O$_3$(0001)/α-Cr$_2$O$_3$(0001) interfaces. In the α-Cr$_2$O$_3$ region, the Cr1$^L$ site becomes the most stable site, and the O4$^U$ and O4$^L$ sites slightly become unstable. However, the effect of the interface on the hydrogen formation energies is insignificant, and is limited to within two atomic layers. On the other hand, in the α-Al$_2$O$_3$ region, the most stable sites of H atoms change from the center of the large interstitial site to the O-adjacent site, resulting in about 1 eV decrease in the hydrogen formation energy. Such stabilization of H atoms remains about 0.5 eV even at the O2$^U$ site, five atomic layers away from the interfaces, while the hydrogen formation energy becomes close to that of the bulk α-Al$_2$O$_3$ at the O1$^U$ sites. Figure 7 shows the differential electron density distribution in the vicinity of H atoms at the O2$^U$ and Cr1$^L$ sites. We can see the regions of increased electron density localized between O-H and Cr-H. We note that the electron density increase between H and O atoms cannot be observed in the bulk α-Al$_2$O$_3$. This result indicates that the O-H covalent bonds can be formed even in the α-Al$_2$O$_3$ region by forming α-Al$_2$O$_3$(0001)/α-Cr$_2$O$_3$(0001) interfaces. This is because the excess electron, resulting from the formation of O-H covalent bonds, can be accepted by $Cr^{3+}$. O-H covalent bonds can be seen even at the $O1^U$ sites. However, although the bonding manner is different, the hydrogen formation energy converges to almost the same value as in the case of the bulk α-$Al_2O_3$. The COHP analysis also shows the existence of O-H covalent bonds. The O-H bond energy in the α-$Al_2O_3$ region is almost constant at about 8 eV.

Although no clear dependence of the bond energy on the distance from the interface was observed, the hydrogen formation energy at O-adjacent sites in the α-$Al_2O_3$ region increased with distance from the interfaces. We evaluate the strain energy induced by H atoms to investigate the origin of the distance dependence of the hydrogen formation energy from the interfaces. The strain energy was defined as the energy difference of the interface model with and without atomic displacement induced by H atoms, as follows:

$$E_s = E_{bulk}^d - E_{bulk}, \qquad (2)$$

where $E_{bulk}^d$ is the total energy of the α-$Al_2O_3$(0001)/α-$Cr_2O_3$(0001) interfaces with atomic displacement induced by H atoms. Figure 8 shows the hydrogen formation energy and strain energy at each O-adjacent site in the α-$Al_2O_3$ region. We can find that O-adjacent sites with higher hydrogen formation energy tend to have higher strain energy. This trend may be because the strain energy induced by H atoms is higher away from the interface, which appears to be softer than the bulk α-$Al_2O_3$ due to non-uniform chemical bonding, with returning to the intrinsic high-strength α-$Al_2O_3$. From these points, the strain energy seems to be one of the significant factors contributing to the dependence of the hydrogen formation energy on the distance from the interfaces.

Finally, we investigated the diffusion property of H atoms at the α-$Al_2O_3$(0001)/α-$Cr_2O_3$(0001) interfaces. Based on the growing process of α-$Al_2O_3$/α-$Cr_2O_3$ composite films, we consider the H atom diffusion from the α-$Al_2O_3$ region to the α-$Cr_2O_3$ region. Figure 9 shows the energy profile of H atom diffusion from the $O1^U$ site to the $O5^L$ site. We can find that the range of hydrogen formation energy through H atom diffusion in the α-$Cr_2O_3$ region of the interface model is almost the same as that in the bulk α-$Cr_2O_3$. Therefore, the diffusion barrier of H atoms in the α-$Cr_2O_3$ region of the interface model, 0.76 eV, is only slightly lower than that in the bulk α-$Cr_2O_3$. In other words, the interfaces do not suppress H atom diffusion in the α-Cr2O3 region.

On the other hand, as discussed before, in the α-$Al_2O_3$ region, the stable sites for hydrogen atoms change to O-adjacent sites, and the hydrogen formation energy is lower than in bulk α-$Al_2O_3$. In contrast to the stable sites, at the saddle point, four atomic layers away from the interfaces, i.e., the Al layer between the O2 and O3 layers, the energy at the saddle point already converges to almost the bulk value. This is because almost all of the valence electrons of Al atoms are transferred to O atoms and cannot form covalent bonds with H atoms even in the vicinity of the interfaces. The lack of Al-H covalent bonds is also evident from the differential electron density distribution shown in Fig. 10. Due to these site-dependent effects of the interface, the H atom diffusion barriers across the O1, O2, and

O3 layers in the α-Al$_2$O$_3$ region are 1.24, 1.47, and 1.19 eV, which are higher than the bulk value. In particular, the H atom diffusion barrier across the O2 layer, which is adjacent to the interfaces, is 0.33 eV higher than the bulk value and has the largest value in the interface model. Finally, we have found that the origin of the high PRF of the α-Al$_2$O$_3$/α-Cr$_2$O$_3$ composite film is the suppression of H atom diffusion by the formation of H atom trapping sites due to the formation of interfaces; however, such H atom trapping sites are not within the interfaces, but rather adjacent to the interfaces.

### 3.3 Isotope effects on hydrogen diffusion

Since D-T fusion reactions consist of deuterium and tritium, we have to consider the isotope effect on hydrogen behavior in TPB. We evaluated the isotope effects on hydrogen formation energy and diffusion barriers by considering the zero-point vibrational energy (ZPE) of hydrogen isotopes. We adopted the harmonic oscillator approximation for the calculation of ZPE. ZPE of H atoms was 0.16 eV at interstitial sites in the α-Al$_2$O$_3$ and 0.30 eV at O-adjacent sites in the α-Cr$_2$O$_3$ and at the α-Al$_2$O$_3$(0001)/α-Cr$_2$O$_3$(0001) interfaces. This is because the H atoms are more strongly confined at the O-adjacent sites through O-H covalent bonds than at the interstitial sites. Since the ZPE of H$_2$ is 0.27 eV, the difference in hydrogen formation energy for H atoms with and without ZPE is about 0.03 and 0.16 eV at the interstitial and O-adjacent sites, respectively. The ZPE of a hydrogen isotope is proportional to the inverse of the square root of its mass. Therefore, the difference in hydrogen formation energies of H, D, and T atoms is less than 0.08 eV in any system, meaning that the isotope effects are very small. In particular, the hydrogen formation energy in the α-Al$_2$O$_3$ is not affected by ZPE, because ZPE at the interstitial sites and in the isolated H$_2$ molecule, which is the energy origin of the hydrogen formation energy, are almost the same. Table 1 shows the diffusion barriers of H, D, and T atoms in the α-Al$_2$O$_3$, the α-Cr$_2$O$_3$, and at the α-Al$_2$O$_3$(0001)/α-Cr$_2$O$_3$(0001) interfaces. For the α-Al$_2$O$_3$(0001)/α-Cr$_2$O$_3$(0001) interfaces, we note that the diffusion barrier is shown when crossing the O2 layer, which has the largest barrier. The diffusion barrier does not change in the α-Al$_2$O$_3$ with ZPE, while it decreases by about 0.1 eV in the α-Cr$_2$O$_3$ and at the α-Al$_2$O$_3$(0001)/α-Cr$_2$O$_3$(0001) interfaces. This trend is because, in the α-Al$_2$O$_3$, the stable site and the saddle point both do not form covalent bonds, while in the α-Cr$_2$O$_3$ and the α-Al$_2$O$_3$(0001)/α-Cr$_2$O$_3$(0001) interfaces, hydrogen spatial confinement is stronger at the O-adjacent sites with O-H covalent bonds than at the saddle points with Cr-H or no covalent bonds. Therefore, there is no isotope effect on the diffusion barrier in the α-Al$_2$O$_3$, while the heavier isotopes in the α-Cr$_2$O$_3$ and at the α-Al$_2$O$_3$(0001)/α-Cr$_2$O$_3$(0001) interfaces have larger diffusion barriers. The difference in diffusion barriers between the α-Al$_2$O$_3$ and the α-Al$_2$O$_3$(0001)/α-Cr$_2$O$_3$(0001) interfaces is 0.21 eV for D atoms and 0.23 eV for T atoms. Even in the cases of heavier isotopes with ZPE, the diffusion barriers are improved by about 20% due to the effect of the interfaces. These results indicate that the hydrogen permeation suppression by the interfaces does not significantly depend on the isotopes.

**3.4 Hydrogen-trapping effects of oxygen vacancies**

Finally, we briefly discuss the effect of oxygen vacancies. It has been well known that the oxygen vacancies show strong hydrogen-trapping effects in the oxide materials. We adopted four oxygen-deficient interface models, introducing a single oxygen vacancy at varying distances from the interface in the α-Al$_2$O$_3$ and α-Cr$_2$O$_3$ regions. In the interface models with oxygen vacancies, the hydrogen impurity is most stable as a hydride at the center of the oxygen vacancy site. Table 2 shows the hydrogen formation energy at the oxygen vacancy sites and the corresponding oxygen vacancy formation energy. The oxygen vacancy formation energy $E_v$ is defined by the following equation:

$$E_v = E_{bulk}^v + \frac{1}{2}\mu_{O_2} - E_{bulk}, \tag{3}$$

where $E_{bulk}^v$ is the total energy of the systems with an oxygen vacancy. $\mu_{O_2}$ is the total energy of molecular oxygen in a cube of 20 Å on a side. The hydrogen formation energy decreased by almost 2.5 eV compared to those without oxygen vacancies. These hydrogen-trapping effects are more significant than the hydrogen diffusion barrier near the interface, indicating that oxygen vacancies act as strong hydrogen-trapping sites. Furthermore, the hydrogen formation energy increases from the α-Cr$_2$O$_3$ region to the α-Al$_2$O$_3$ region, similar to the cases without oxygen vacancies, and no significant stabilization of hydrogen impurities at the interface sites was observed. The oxygen vacancy formation energy also increases monotonically from the α-Cr$_2$O$_3$ region to the α-Al$_2$O$_3$ region. Compared with bulk α-Al$_2$O$_3$ and α-Cr$_2$O$_3$, as reported in [46], the oxygen vacancy formation energy obtained from the interface models is approximately 1 eV lower. Therefore, while the formation of oxygen vacancies is slightly enhanced in the vicinity of the interface, the interface site had no specific effect on the formation of oxygen vacancies. We also noted that although oxygen vacancies near the interface are easier to form than in bulk α-Al$_2$O$_3$ and α-Cr$_2$O$_3$, corresponding formation energy still remains very high, exceeding 4.71 eV.

**4. Conclusion**

In order to reveal the effects of the composite layering of TPB on the diffusion behavior of hydrogen isotopes, we have investigated the hydrogen formation energy and the diffusion barriers of hydrogen isotopes in the α-Al$_2$O$_3$, the α-Cr$_2$O$_3$, and at the α-Al$_2$O$_3$(0001)/α-Cr$_2$O$_3$(0001) interfaces using *ab initio* electronic structure calculations based on DFT. Hydrogen isotopes do not form covalent bonds with surrounding atoms throughout the diffusion path in the α-Al$_2$O$_3$, which results in the large hydrogen formation energy and diffusion barrier. On the other hand, hydrogen isotopes form O-H covalent bonds at the most stable sites and Cr-H covalent bonds at the saddle point in the α-Cr$_2$O$_3$. Therefore, the α-Al$_2$O$_3$ is a more promising candidate material as TPB than the α-Cr$_2$O$_3$. In the interface model, the diffusion properties of hydrogen isotopes and the effects of the interfaces differ between the α-Al$_2$O$_3$ and the α-Cr$_2$O$_3$ regions. In the α-Cr$_2$O$_3$ region, the interfaces have only a small effect on the diffusion behavior of hydrogen isotopes, while in the α-Al$_2$O$_3$ region, they even affect the bonding manner of hydrogen isotopes. With the formation of the interfaces, in the α-Al$_2$O$_3$ region, the most stable sites for hydrogen isotopes change from the center of the largest interstitial site to O-adjacent sites by forming O-H covalent bonds. On the other hand, no covalent bond between hydrogen isotopes and surrounding atoms still formed at the saddle points. Therefore, the diffusion barriers at the α-Al$_2$O$_3$(0001)/α-Cr$_2$O$_3$(0001) interfaces become about 0.2 eV larger than those in the bulk α-Al$_2$O$_3$. The hydrogen isotope trapping sites that contribute the most to suppressing hydrogen isotope diffusion are not the interface sites but the sites adjacent to the interfaces. Furthermore, it was revealed that, similar to bulk α-Al$_2$O$_3$ and α-Cr$_2$O$_3$, oxygen vacancies act as hydrogen-trapping sites. While the formation energy of oxygen vacancies decreased by approximately 1 eV through the formation of the interfaces, no significant increase in the formation of oxygen vacancies or the hydrogen-trapping effect was observed at the interface sites. These results shed light on the mechanism by which α-Al$_2$O$_3$/α-Cr$_2$O$_3$ composite films suppress hydrogen isotope diffusion and why they are promising as TPBs. The further detailed quantitative discussion of whether vacancies play a dominant role in hydrogen permeation suppression requires additional studies, including the various charge states of vacancies. In this study, we treated interfaces composed of the (0001) plane orientation to model the epitaxial growth. In the future, studying various plane orientations and ceramic materials will provide useful insight into developing new composite films for more efficient TPB. In addition, it is required to elucidate the effects of defects formed as irradiation damage, such as complex of atomic vacancies and interstitial atoms, dislocation loops, and cavities, as these may be strong trapping sites.

**Declaration of competing interest**

The authors declare that they have no competing financial interests or personal relationships that could have appeared to influence the work reported in this paper.

**Acknowledgments**

This work has been done using the facilities of the Supercomputer Center, the Institute for Solid State Physics, the University of Tokyo.

Table 1 Diffusion barriers of hydrogen isotopes (H, D, T) in the α-Al$_2$O$_3$, in the α-Cr$_2$O$_3$, and at the α-Al$_2$O$_3$(0001)/α-Cr$_2$O$_3$(0001) interfaces.

|  | Diffusion barrier (eV) | | | |
| --- | --- | --- | --- | --- |
|  | H | D | T | without ZPE |
| α-Al$_2$O$_3$ | 1.15 | 1.15 | 1.15 | 1.14 |
| α-Cr$_2$O$_3$ | 0.79 | 0.83 | 0.85 | 0.92 |
| α-Al$_2$O$_3$(0001)/α-Cr$_2$O$_3$(0001) | 1.31 | 1.36 | 1.38 | 1.47 |

Table 2 Hydrogen formation energy $E_f$ at the oxygen vacancy sites and oxygen vacancy formation energy $E_v$ near the α-Al$_2$O$_3$(0001)/α-Cr$_2$O$_3$(0001) interfaces.

| Region | Distance from the interface (Å) | $E_f$ (eV) | $E_v$ (eV) |
| --- | --- | --- | --- |
| α-Al$_2$O$_3$ | 7.632 | 0.12 | 6.80 |
| | 1.125 | −0.26 | 6.16 |
| α-Cr$_2$O$_3$ | 1.125 | −0.40 | 4.90 |
| | 5.817 | −0.46 | 4.71 |

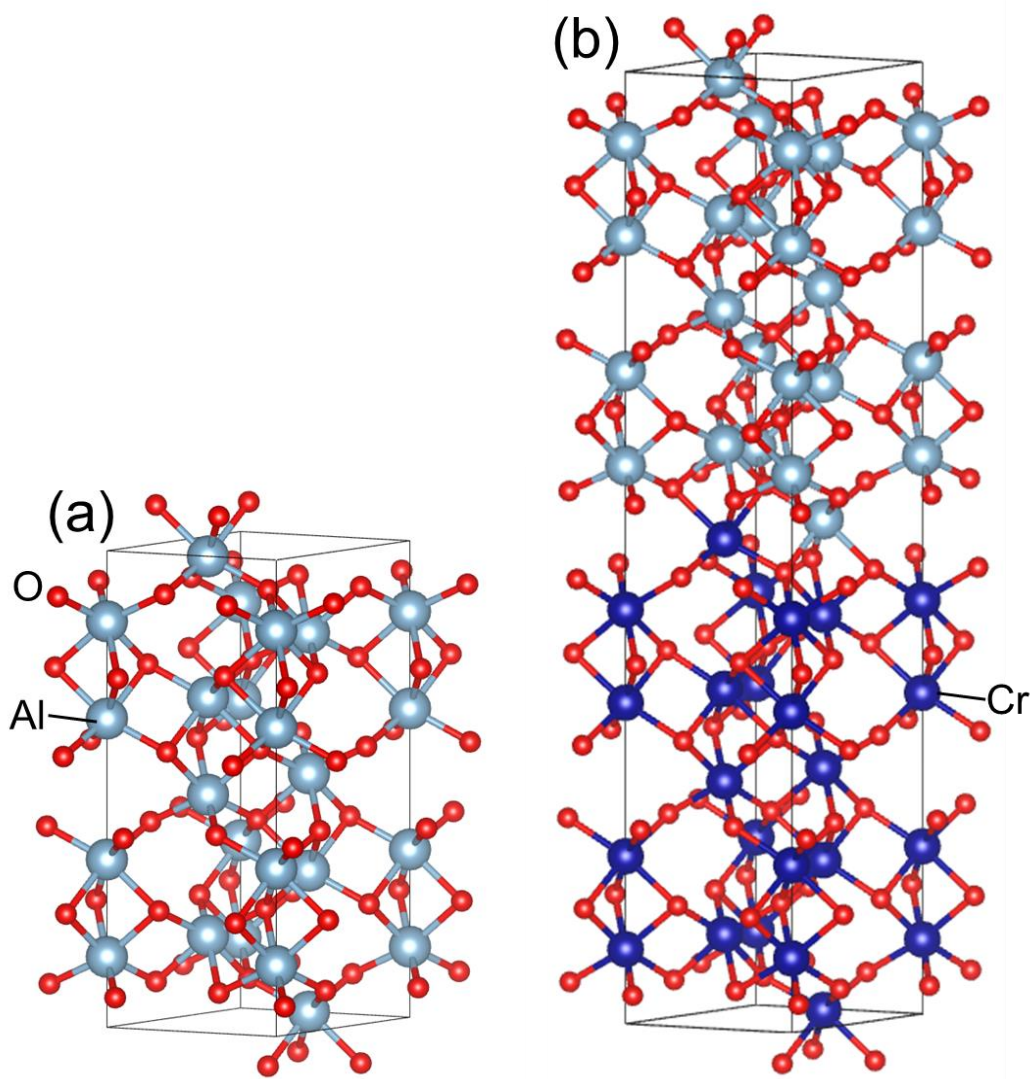

Fig. 1 Supercell for (a) α-Al$_2$O$_3$, and (b) α-Al$_2$O$_3$(0001)/α-Cr$_2$O$_3$(0001) interfaces. The light blue, dark blue, and red balls represent Al, Cr, and O atoms.

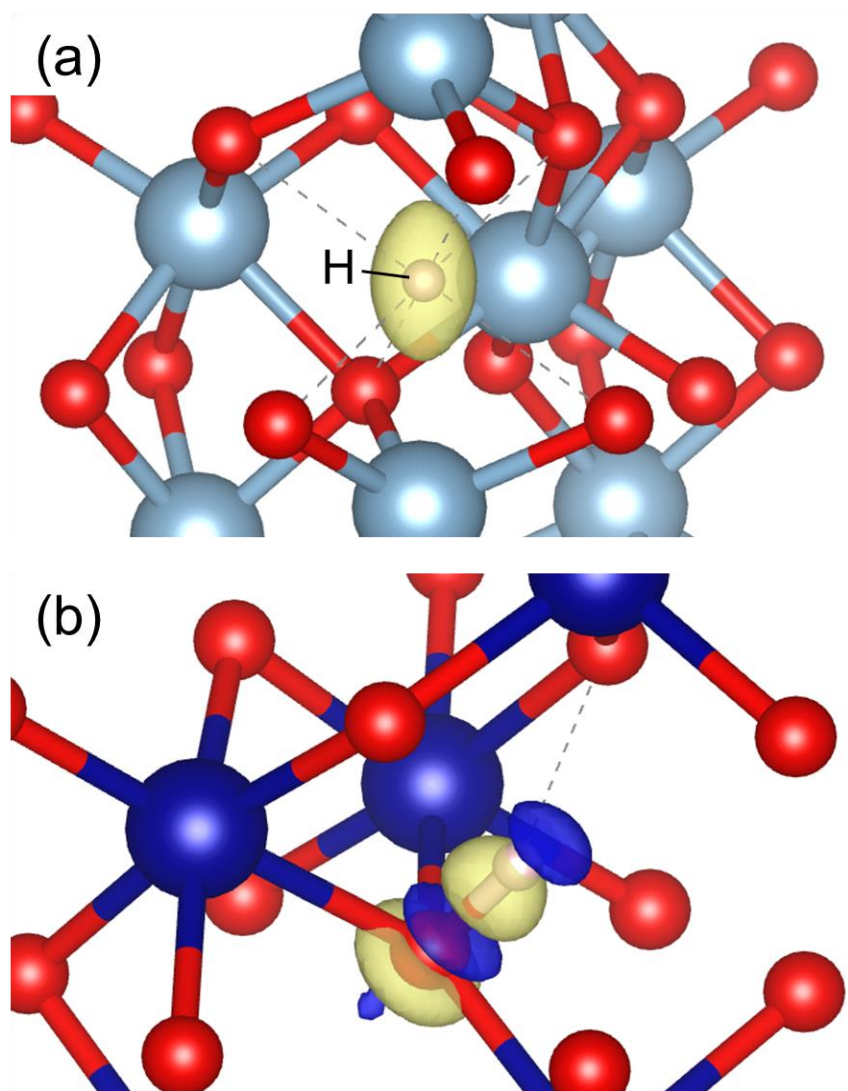

Fig. 2 Differential electron density distribution in the vicinity of H atoms in the (a) α-Al$_2$O$_3$ and (b) α-Cr$_2$O$_3$. The light blue, dark blue, red, and white balls represent Al, Cr, O, and H atoms. Yellow (lighter) and blue (darker) regions show the regions where electron density increases and decreases. Isosurface values are set to $\pm 0.1\ e/\text{Å}^3$.

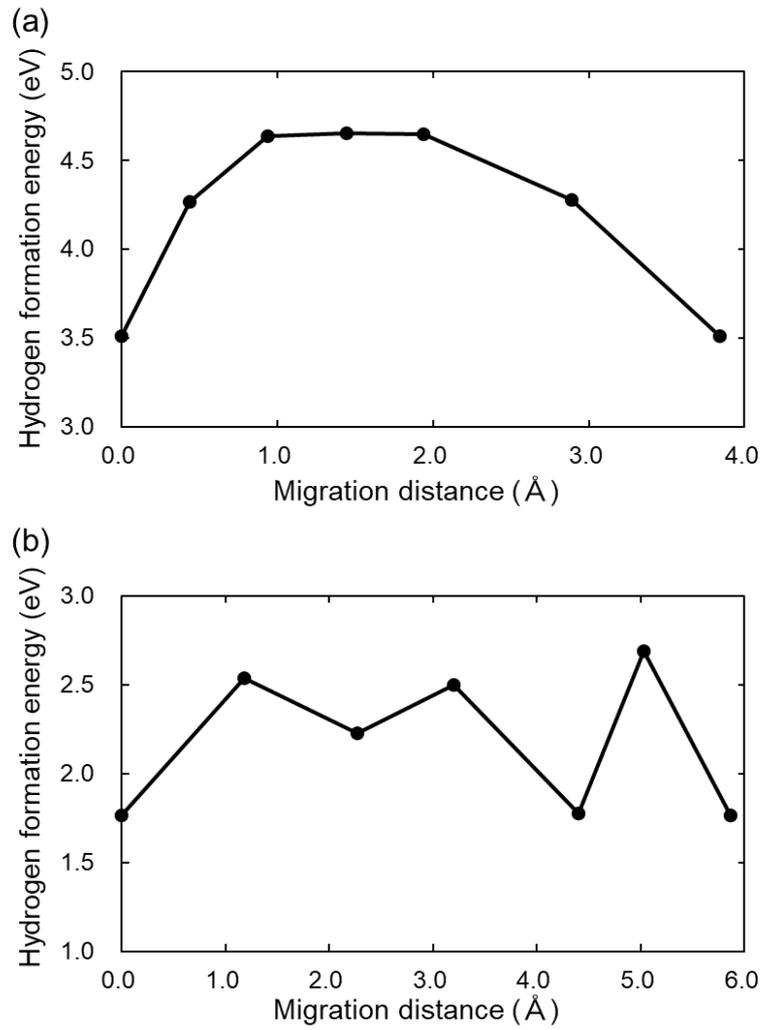

Fig. 3 Energy profiles through H atom diffusion between the most stable sites in the (a) α-Al$_2$O$_3$ and (b) α-Cr$_2$O$_3$.

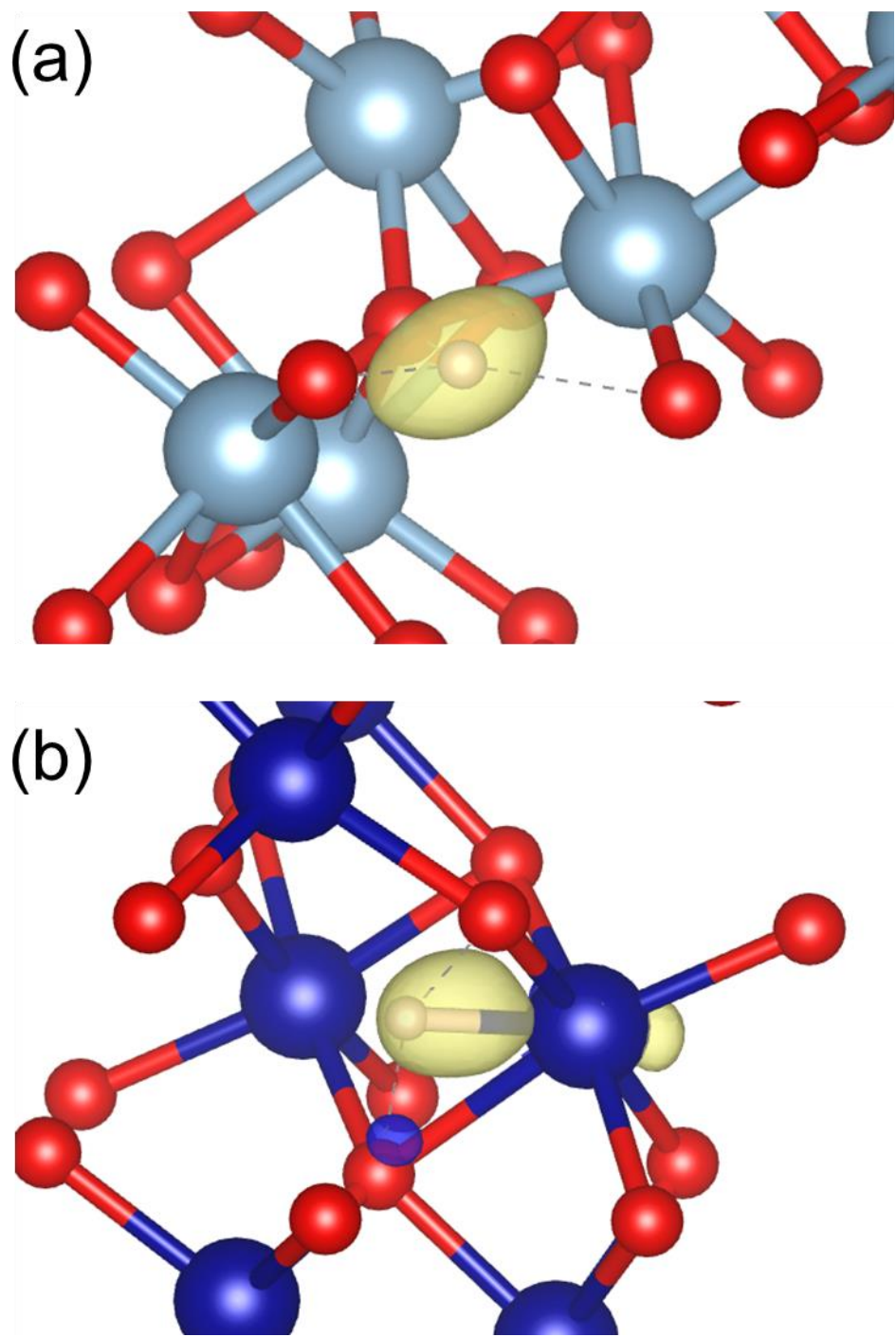

Fig. 4 Differential electron density distribution at the saddle points of H atom diffusion in the (a) α-Al$_2$O$_3$ and (b) α-Cr$_2$O$_3$. The light blue, dark blue, red, and white balls represent Al, Cr, O, and H atoms. Yellow (lighter) and blue (darker) regions show the regions where electron density increases and decreases. Isosurface values are set to $\pm 0.1$ $e/\text{Å}^3$.

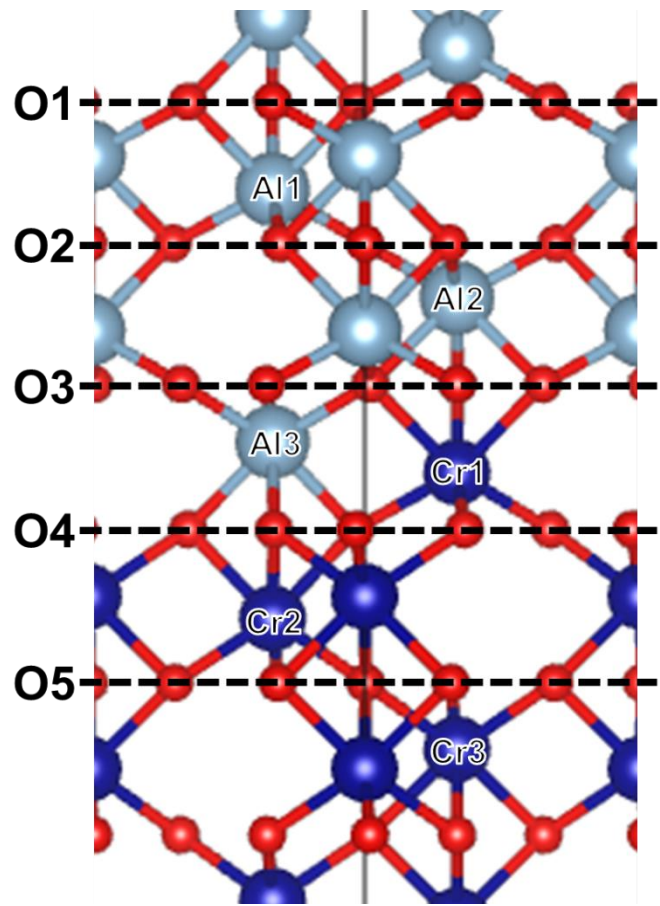
Fig. 5 Initial sites for H atoms at the α-Al$_2$O$_3$(0001)/α-Cr$_2$O$_3$(0001) interfaces.

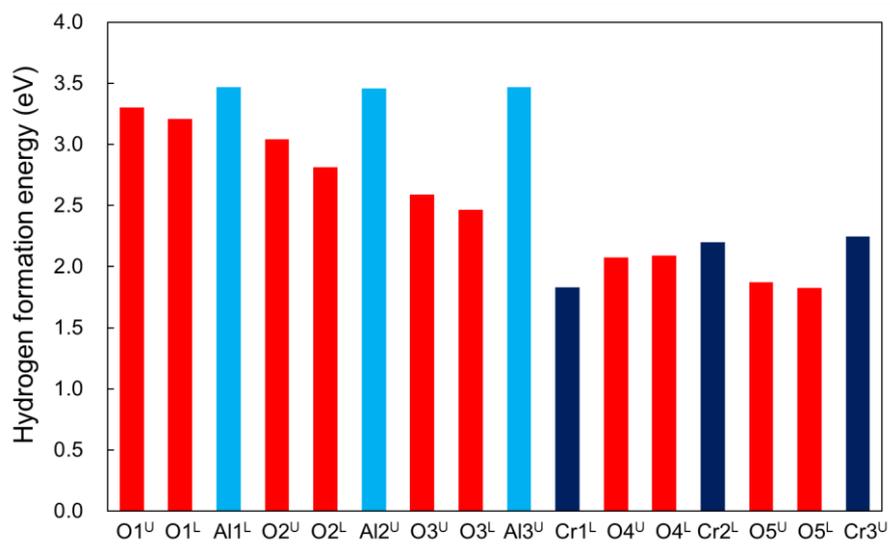

Fig. 6 Hydrogen formation energy at α-$Al_2O_3$(0001)/α-$Cr_2O_3$(0001) interfaces.

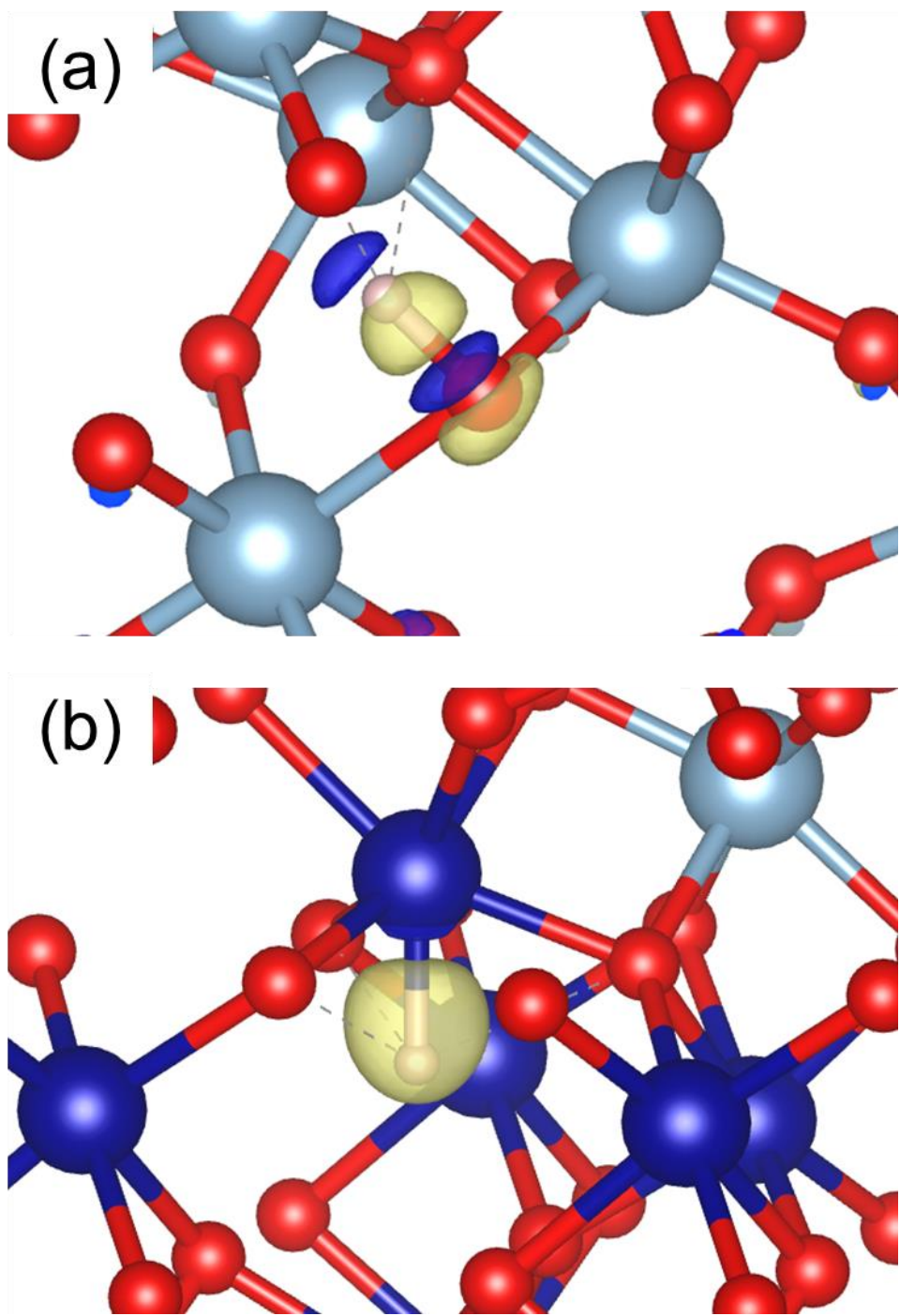

Fig. 7 Differential electron density distribution in the vicinity of H atoms at the (a) $O2^U$ and (b) $Cr1^L$ sites. The light blue, dark blue, red, and white balls represent Al, Cr, O, and H atoms. Yellow (lighter) and blue (darker) regions show the regions where electron density increases and decreases. Isosurface values are set to $\pm 0.1\ e/\text{Å}^3$.

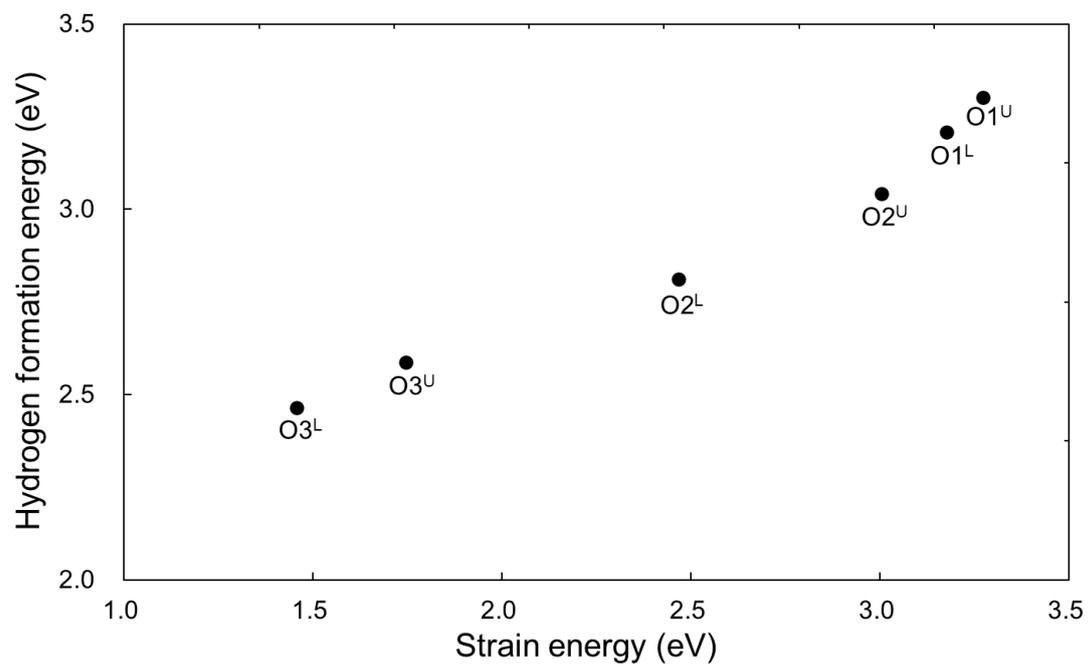

Fig. 8 Correlation between solid solution energy of H atoms and strain energy in the α-Al2O3 region near the interface.

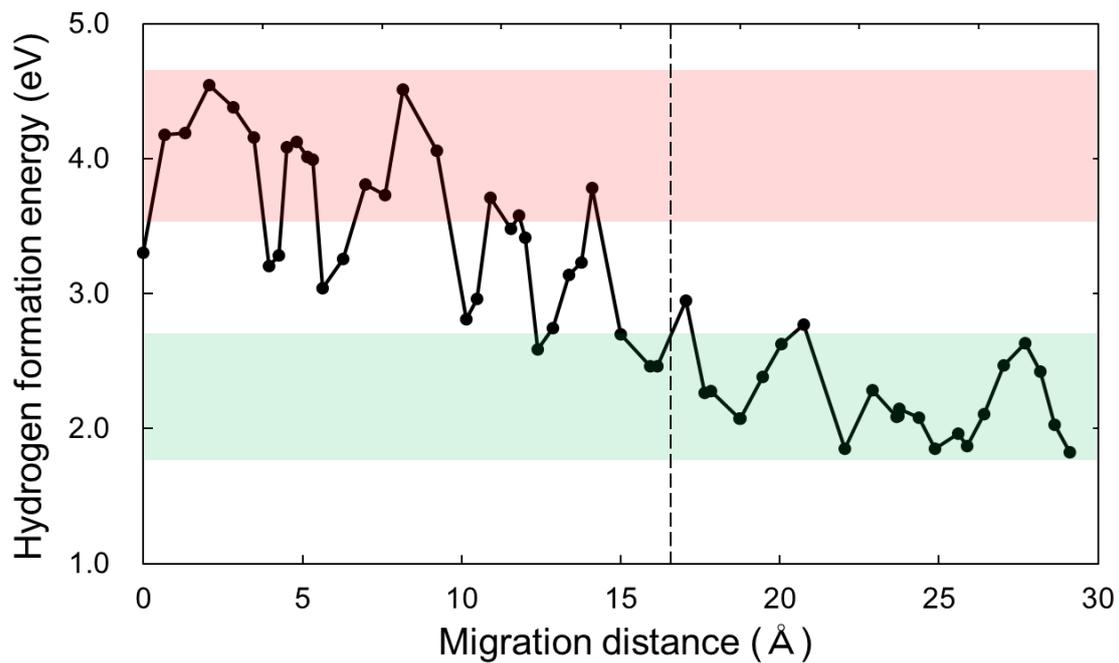

Fig. 9 Energy profiles through H atoms diffusion at α-Al$_2$O$_3$(0001)/α-Cr$_2$O$_3$(0001) interfaces. The dotted line indicates the position of the interfaces, with the α-Al$_2$O$_3$ region on the left and the α-Cr$_2$O$_3$ region on the right. The pink (upper) and green (lower) bands indicate the range of hydrogen formation energy from the most stable sites to saddle points in the bulk α-Al$_2$O$_3$ and α-Cr$_2$O$_3$, respectively.

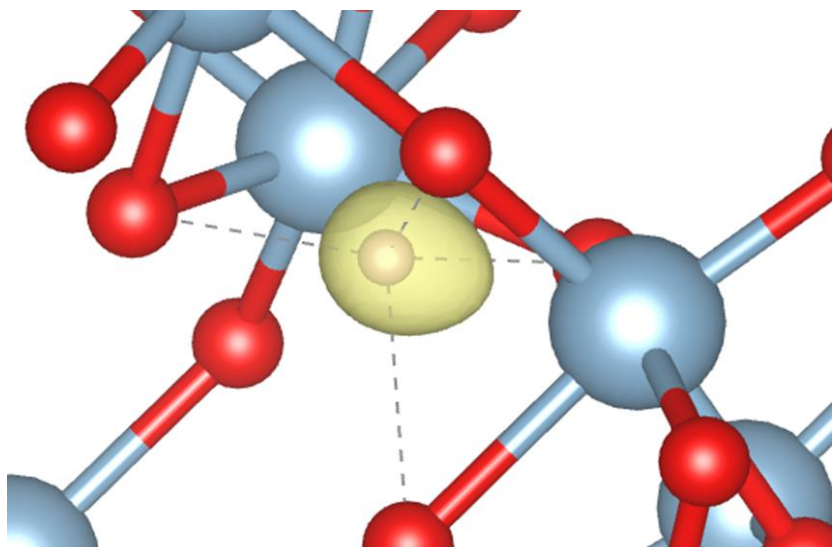

Fig. 10 Differential electron density distribution in the vicinity of H atoms at the saddle point in the O2 layer. The light blue, red, and white balls represent Al, O, and H atoms. Yellow (lighter) and blue (darker) regions show the regions where electron density increases and decreases. Isosurface values are set to $\pm 0.1\ e/\text{Å}^3$.